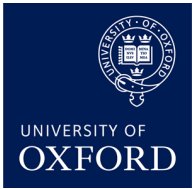
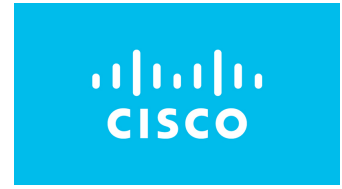

# Ethics and Responsible AI Deployment

Dr Petar Radanliev[1,2], Omar Santos[3]

[1]Department of Computer Sciences, University of Oxford

`[`[2]Department of Engineering Sciences, University of Oxford

[3]Cisco Systems, RTP, North Carolina, United States

Correspondence email: <petar.radanliev@cs.ox.ac.uk>

**Abstract**: As Artificial Intelligence (AI) becomes more prevalent, protecting personal privacy is a critical ethical issue that must be addressed. This article explores the need for ethical AI systems that safeguard individual privacy while complying with ethical standards. By taking a multidisciplinary approach, the research examines innovative algorithmic techniques such as differential privacy, homomorphic encryption, federated learning, international regulatory frameworks, and ethical guidelines. The study concludes that these algorithms effectively enhance privacy protection while balancing the utility of AI with the need to protect personal data. The article emphasises the importance of a comprehensive approach that combines technological innovation with ethical and regulatory strategies to harness the power of AI in a way that respects and protects individual privacy.

## 1. Introduction: The Role of Algorithms in Protecting Privacy

In today's world, as Artificial Intelligence (AI) continues to develop and expand, ethical concerns have become crucial in discussions surrounding its deployment. One of the most significant concerns is the protection of privacy [1]. As AI systems become more prevalent in various aspects of society, handling enormous amounts of data, including sensitive and personal information, the potential for privacy infringement rises [2]–[8]. Therefore, it is necessary to use AI algorithms that not only respect but also actively safeguard privacy.

The challenge of aligning AI with ethical norms, particularly in privacy protection, is multifaceted. It involves balancing utilising AI's capabilities for societal benefit while ensuring that individual privacy rights are not compromised. This balancing act is further complicated by the inherent complexities in AI technologies, which often operate as 'black boxes,' making it difficult to understand how data is being processed, and conclusions are being drawn.

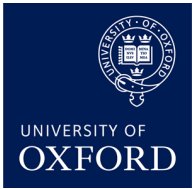
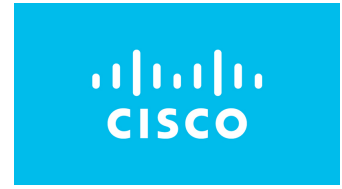

Algorithms play a critical role in addressing these challenges [9]–[19]. When designed with privacy protection as a core principle, algorithms can act as guardians of personal data. Techniques such as differential privacy, homomorphic encryption, and federated learning have emerged as powerful tools. Differential privacy allows for extracting valuable insights from datasets while mathematically guaranteeing the anonymity of individual data points. Homomorphic encryption enables data to be processed in its encrypted form, thus offering a new paradigm for secure data analysis. Federated learning allows AI models to be trained across multiple decentralised devices holding local data samples without exchanging them, thereby preserving privacy.

Further, developing AI systems that prioritise privacy requires a collaborative approach involving technologists, ethicists, policymakers, and the broader community. Establishing international standards and ethical guidelines and ensuring compliance with regulatory frameworks are essential steps in this journey. Only through such a comprehensive and multidisciplinary approach can AI be harnessed as a tool for enhancing privacy rather than diminishing it.

Therefore, this article aims to explore the intricate relationship between AI and privacy, focusing on how algorithms can be leveraged to protect and enhance privacy in the age of ubiquitous AI. It seeks to provide a nuanced understanding of the ethical challenges involved in AI deployment and how they can be addressed through innovative algorithmic solutions and collaborative efforts.

## 2. Case Study of the Bletchley Summit

On the first day of the AI safety summit, the UK, US, EU, Australia, and China signed the Bletchley Declaration - the first international agreement to deal with the rapidly emerging technology, artificial intelligence (AI). The 28 governments that signed the declaration acknowledged that AI poses a potentially catastrophic risk to humanity. They agreed to collaborate on AI safety research and work together to regulate AI development, despite signs of competition between the US and the UK to take the lead in developing new regulations.

Michelle Donelan, the UK technology secretary, emphasised the importance of looking not just independently, but collectively at the risks around frontier AI. Frontier AI refers to the most cutting-edge systems, which some experts believe could become more intelligent than people at various tasks.

Elon Musk, the owner of Tesla and SpaceX, and of X, formerly Twitter, warned that this is the first time we have a situation where there's something that is going to be far smarter than the smartest human. He added that it's not clear if we can control such a thing.

The declaration included the sentence, "We welcome the international community's efforts so far to cooperate on AI to promote inclusive economic growth, sustainable development, and innovation, to protect human rights and fundamental freedoms, and to foster public trust and confidence in AI systems to realise their potential fully."

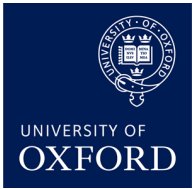
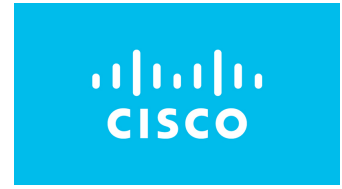

Wu, one of the Chinese delegates, told fellow delegates that they uphold the principles of mutual respect, equality, and mutual benefits. Regardless of their size and scale, countries have equal rights to develop and use AI.

However, there currently needs to be more international agreement on what a global set of AI regulations might look like or who should draw them up. Some British officials had hoped other countries would agree to strengthen the government's AI task force to test new models from around the world before they are released to the public. But instead, Raimondo used the summit to announce a separate American AI Safety Institute within the country's National Institute of Standards and Technology. She called it "a neutral third party to develop best-in-class standards" and added that the institute would develop its own safety, security, and testing rules. Kamala Harris, the vice president, then gave a speech on AI in London in which she talked about the importance of regulating existing AI models and more advanced ones in the future.

During a recent interview, UK minister Amanda Solloway and US Commerce Secretary Gina Raimondo praised the partnership between the UK and the US safety institute, dismissing any suggestion of a split between the two countries on which country should take the global lead on AI regulation. Meanwhile, the EU is passing an AI bill to develop a set of principles for regulation and bring in rules for specific technologies, such as live facial recognition. However, UK minister Michelle Donelan has suggested that the government would not include an AI bill in the king's speech next week, as they need to properly understand the problem before applying the solutions.

But she denied the UK was falling behind its international counterparts.

## 3. Ethical considerations in AI decision-making

Incorporating artificial intelligence (AI) into various spheres of life has led to various ethical considerations that necessitate delicate handling. The topic of "Ethical considerations in AI decision-making" [20] covers various issues requiring careful consideration [1], [21]. These issues include but are not limited to, the accountability and transparency of AI systems, the ethical implications of AI on privacy and security, the potential for AI to perpetuate existing biases and discrimination, and the responsibility of developers and users to ensure that AI is employed ethically and responsibly. Addressing these concerns is paramount, as it will ensure that the benefits of AI are maximised while minimising its potential negative impacts. By adopting an ethical framework that promotes transparency, accountability, and fairness, we can ensure that AI is employed ethically and in the best interests of society.

### 3.1. How Can Algorithms Help to Protect our Privacy

Data protection is critical to modern security, and encryption algorithms play a significant role. Their primary goal is to convert data into an indecipherable format without a specific key. Using mathematical operations, encryption techniques transform plaintext data into ciphertext, making it appear gibberish to anyone without the proper decryption key [22]. Encryption has many practical uses, including securing online communications, safeguarding sensitive information in databases, and protecting files on electronic devices [23].

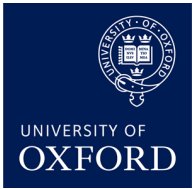
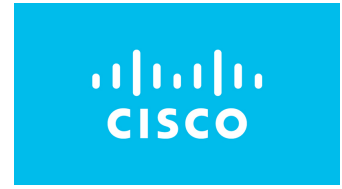

Homomorphic Encryption is a ground-breaking encryption technique that allows computations on encrypted data without needing to access the plaintext. This approach ensures that all operations are carried out on the encrypted data itself, with only the result being decrypted to reveal significant information. Homomorphic Encryption has many applications, including encrypted search, secure cloud computing, and privacy-preserving computations in medical research.

Differential Privacy is a method that protects individual data points by ensuring that the addition or deletion of a single data point has no discernible impact on the results of any study. This is achieved by obscuring individual data entries with noise that is put into the data or queries conducted on the data. Differential Privacy is useful in preserving user information in analytics, maintaining privacy in machine learning datasets, and protecting user data in census databases.

Secure Multi-Party Computation (SMPC) is a technique that enables private computation of a function over the inputs of multiple participants. The data is divided into several sections and dispersed to different parties. These distinct components are subjected to computations that keep the original data hidden. SMPC has many applications, including safe voting procedures, collaborative data-driven projects without data sharing, and privacy-preserving medical research.

In summary of this section, encryption techniques like Homomorphic Encryption, Differential Privacy, and SMPC play a vital role in modern data security. They provide a secure and robust way to protect sensitive information, ensure privacy, and enable secure computations on encrypted data [24].

### 3.2. **Other** Ethical considerations in AI decision-making

The rules concerning AI accountability in the legal and regulatory domain constantly evolve. In this regard, GDPR [25], [26] has significant implications for AI accountability. Those responsible for creating and using AI decision-making systems must take responsibility for the choices made by the system. This includes monitoring the system for biases and errors and taking the necessary steps to minimise adverse effects.

AI systems must also respect individual privacy and adhere to strict data protection guidelines to ensure that personal information is not compromised. Addressing bias in AI algorithms and prioritising fairness to eliminate discriminatory outcomes in AI decision-making processes is crucial. Maintaining accountability and transparency in AI decision-making is essential. Therefore, it is necessary to examine accountability in artificial intelligence, from model creators to consumers.

The ethical implications of AI-driven decision-making in military applications are also highlighted in the AI in Warfare and Defence [27]. Specifically, the challenges of moral responsibility and autonomous weaponry are discussed.

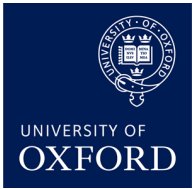
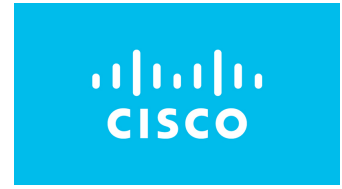

# 4. Addressing bias, transparency, and accountability

In today's world, AI systems play an increasingly significant role in decision-making processes. However, if the data used to train these systems is biased or skewed, it can lead to accurate or fair outcomes. Therefore, it's crucial to analyse the effects of such data on AI models and develop methods to reduce bias. In this regard, this article delves into an in-depth analysis of how skewed datasets affect AI judgement. It provides strategies for reducing bias in AI, from gathering data to assessing models.

## 4.1. Techniques for Ensuring Transparency

Transparency is vital to AI models, especially complex ones like deep neural networks. This is essential for improving the model, addressing ethical concerns, and gaining broader acceptance in critical industries. Feature significance methods are crucial in achieving transparency by identifying the input features pivotal to the model's predictions.

### 4.1.1. Feature Importance Methods

Feature importance is a crucial metric that assesses the significance of each input feature concerning the predictive capacity of a model. This assists stakeholders in comprehending the model's behaviour and detecting potential biases. Moreover, pinpointing the features that hold more weight can enhance and refine the model's performance while streamlining its complexity by eliminating extraneous elements.

*Permutation importance*

Permutation importance is a useful technique that helps to determine how important individual features are in relation to a machine learning model's predictive capability. The process involves randomly shuffling the values of a specific feature and observing the resulting change in the model's performance. If mixing a feature's values significantly drops the model's performance, the feature is essential. Conversely, if shuffling has little or no effect, the feature may not be crucial for the model's predictions. This approach can be a valuable tool for identifying key features in a model and improving its overall performance.

Permutation importance is a method used to determine the importance of each feature in a machine-learning model. This process involves several steps that help establish a performance baseline, disrupt the relationship between a selected element and the target variable, assess how much the feature disruption affects performance, and calculate the feature's relevance based on the observed performance loss.

The first step in the permutation importance process is to record the model's performance with the original dataset. This step establishes a performance baseline that can be used to compare subsequent evaluations. After training the model on the original dataset, its performance is measured using a chosen metric such as accuracy, F1-score, Mean Absolute Error, or any other relevant metric depending on the problem type (classification, regression, etc.).

The second step involves shuffling the values of a particular feature across the dataset. This step disrupts the relationship between the selected feature and the target variable by randomly permuting the values of a single feature column. This process ensures that the distribution of the feature remains the same, but any correlation or relationship it may have with the output is destroyed. It is important to keep the other feature columns unchanged.

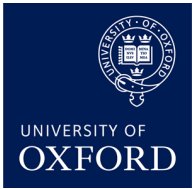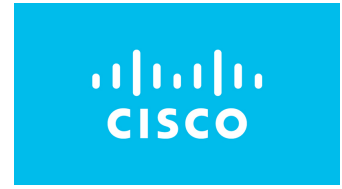

The third step entails reassessing the model's effectiveness using the permuted data. This step aims to calculate how much the disruption of the selected feature affects performance. For the dataset with the permuted feature column, predict using the pre-trained model. Then, use the same metric from step one to assess the model's performance. Importantly, we're only using the model to forecast on modified data—we're not retraining it.

The final step involves calculating the feature's importance. We can calculate a numerical number representing the feature's relevance based on the observed performance loss. To calculate the importance of a feature, we deduct the performance metric found in step three from the initial step metric (step one). When a feature's disruption substantially impacts the model's predictions, a larger drop indicates the feature's high importance. A tiny drop suggests that the attribute may not be very important for the model to choose.

However, there are some important pointers to remember while using the permutation importance method. Permutation significance can shed light on the relative relevance of different characteristics, but it may not be a good way to record feature interactions. If two traits have a strong correlation or interaction, permuting one might not significantly affect performance if the other is left unchanged. The process should be done several times (using different random permutations) to acquire a reliable estimate of importance, particularly when data may be naturally variable.

Comparing a model's characteristics allows us to rank them according to their significance, which facilitates feature selection, simplifies the model, and improves interpretability. By understanding the relative importance of each feature, we can modify the model to improve its performance and better understand how it makes predictions.

Permutation importance is a powerful and effective approach widely used to evaluate the importance of features in a machine learning model. It is a model-agnostic technique that systematically modifies features and observes the resulting performance degradation. It provides valuable insights into the model's behaviour, even in a complex "black-box" model [28]. This approach promotes transparency and boosts confidence in AI systems, making it an essential tool for machine learning experts and practitioners.

One of the key benefits of permutation importance is that it applies to any model, regardless of its complexity or structure. This means that it can be used to evaluate the importance of features in a wide range of machine learning models, including decision trees, neural networks, and support vector machines. Moreover, after multiple trials, the importance of permutation consistently produces reliable outcomes, making it a robust and reliable technique for feature selection and model optimisation.

However, some things could be improved in using permutation importance. One of the main challenges is the computational complexity associated with evaluating the performance of each feature multiple times. This can be time-consuming and resource-intensive, especially when dealing with large datasets or complex models. Additionally, permutation importance does not consider feature interactions, which can be an important factor in some applications. Despite these limitations, permutation is a widely used and valuable machine learning feature selection and model evaluation technique.

*Drop-column importance*

In machine learning interpretability, drop-column importance is an evaluation technique that measures the contribution of a specific feature to a model's prediction performance. This approach involves comparing the model's performance with all its elements to its performance with a particular feature removed. By retraining the model without the feature and assessing its performance, we can quantify the significance of the deleted feature and gain insights into its relevance.

Drop-column importance is a valuable tool for feature selection and model transparency. It helps us identify redundant features that have little impact on the model's performance when removed. On the other hand, features that significantly impair performance when removed are considered essential to the model's predictions.

The process of using this technique involves two main steps:

1. Determine Baseline Performance:

The first step is understanding the model's potential performance using the entire dataset. This involves training the model with all features and evaluating its output using an appropriate metric such as accuracy, F1-score, mean squared error, or any other relevant statistic, depending on the problem. This step provides a benchmark performance measure that illustrates the model's potential.

2. Feature Extraction:

   The second step examines how the model is affected when removing a feature. This involves removing a single feature from the dataset, altering the dataset, and losing all entries related to the selected element. By retraining the model without the feature and assessing its performance, we can determine its importance and gain insights into its relevance.

Drop-column importance is a powerful technique that can help us select the most relevant features for our model and increase its transparency.

Assessing and retraining a machine learning model involves determining the model's performance when a specific feature is removed. To accomplish this, the first step is to train the model using an updated dataset, but this time, without the feature that needs to be evaluated. The same metric used in the previous step is then employed to gauge the model's performance. The outcome of this step is a performance metric that can be compared to the baseline to determine the feature's importance.

Evaluating feature significance involves determining the importance of a feature by subtracting the performance measure (with the feature removed) from the initial performance measure produced in the previous step. If the drop in performance is significant, the feature was essential to the model; otherwise, it may be redundant or of lesser importance. The outcome of this step is a quantitative or qualitative assessment of the feature's significance.

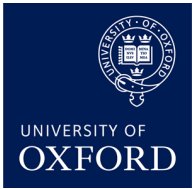
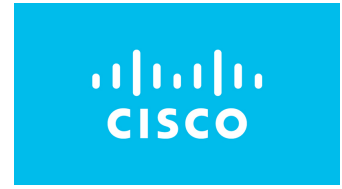

One of the significant advantages of this approach is that it is simple and easy to employ for any machine-learning model. However, the process can be computationally demanding, particularly for datasets with many features, as the model must be retrained for each feature. Moreover, the method may need to account for complex relationships between features as each is assessed separately.

Evaluating feature significance immediately assesses the impact of removing a feature and is ideal for models with interacting features. However, it is more computationally intensive than permutation importance since training the model is required for each feature. Additionally, this method may not be suitable for models with high-dimensional input spaces.

*Model-specific methods*

Machine learning interpretability strategies can be classified into model-specific and generalisable [2], [13], [29]–[47]. Generalisable techniques are designed to work with different model types. In contrast, model-specific methods rely on specific models' unique characteristics and architecture to provide insights into how they make predictions.

One example of a model-specific approach is coefficient analysis, used in linear models such as linear, logistic, and ridge/lasso regression. Linear models have a coefficient for each feature, where the coefficient of a feature represents the expected change in output for a one-unit change in the feature, assuming all other features remain constant.

By analysing the magnitude and sign of these coefficients, we can determine the significance and direction of the relationship between each feature and the target variable. Larger absolute coefficient values indicate greater significance, while the sign indicates the relationship's direction, whether positive or negative.

This method is particularly useful in datasets with many features, as it can help identify correlations and feature relevance. Additionally, regularisation algorithms like Lasso can effectively execute feature selection by reducing some coefficients to zero.

Another topic to consider is related to the methods for tree visualisation in random forests and decision trees. These models effectively segment the feature space to make accurate predictions. It's worth noting that random forests are made up of decision trees that work together, which makes them even more powerful.

In terms of definition, a decision tree works by recursively dividing the data based on feature values until a decision is reached. The tree's branches visually represent the decision-making process, which can be quite insightful.

One of the benefits of decision trees is their interpretability. They allow users to gain valuable insights into the importance of specific features when making predictions. Following the path from the root to the leaves makes it possible to understand how a decision was made. Additionally, the depth at which a feature is used to split the data indicates its significance, with deeper splits showing greater importance.

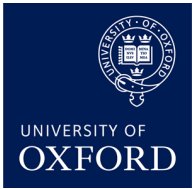
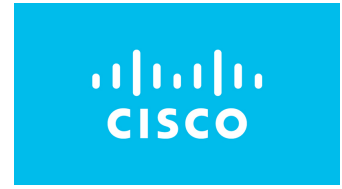

Regarding applications, tree visualisation can be incredibly useful for identifying important features and understanding complex decision-making processes. It's also possible to determine the overall relevance of each feature in a Random Forest by summing up the feature importance across all trees.

*Post-hoc interpretation methods*

Post-hoc interpretation approaches are useful for understanding and interpreting machine learning models after they have been developed. Unlike methods specific to a model, post-hoc techniques are often model-agnostic, meaning they can clarify the decision-making processes of a wide range of models.

One popular post-hoc interpretation method is SHapley Additive exPlanations (SHAP). SHAP values provide a consistent way to assess the importance of features for any machine learning model by assigning importance values to each feature for a particular prediction.

SHAP is based on the Shapley values concept from game theory, which distributes a payout among players in a cooperative game. SHAP uses this idea to distribute each feature's contribution to a specific prediction. SHAP values have two key characteristics. First, they ensure local accuracy by making the sum of SHAP values of all features for each instance the difference between the model's forecast and the baseline prediction. Secondly, SHAP values ensure consistency by assigning a higher value to a feature that a model depends on more for a prediction than it does on another feature.

SHAP values are helpful for both local and global models, providing insights into overall feature relevance and specific prediction explanations. It is particularly useful for complex models where traditional methods of feature relevance determination may need to be revised.

The technique known as Layer-wise Relevance Propagation (LRP) is mainly used to understand deep neural networks, particularly convolutional neural networks (CNNs) [48]. LRP distributes the neural network's prediction backwards via the layers to function. For every neuron, it calculates a relevance score that indicates how much the neuron contributed to the final prediction.

CNNs can be challenging to interpret, but LRP (Layer-wise Relevance Propagation) provides a way to visualise the relevance values and generate a heatmap that shows the regions of input data, such as an image, that are most important to the model's prediction.

LRP is particularly useful in computer vision, where understanding the specific components of an image that lead to a certain classification can provide insight into how the model works. Additionally, it can help identify areas of misfocus, which can aid in debugging and improve overall model performance.

*Model-agnostic methods*

Artificial intelligence models can be complex and challenging to understand because they rely on vast amounts of data and intricate algorithms to make predictions. This can make determining how the model arrived at a specific decision difficult.

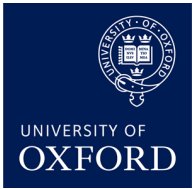
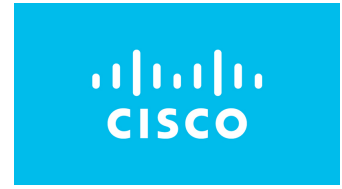

To provide clarity, model-agnostic techniques can be used with almost any type of model, including deep neural networks and linear regressors. These methods are flexible and aim to offer insights into a model's prediction processes regardless of the structures or algorithms used.

Model-agnostic interpretation methods play a vital role in enhancing the transparency and credibility of AI systems by providing valuable insights into how the model operates. They help explain the prediction process and identify the features that contribute most to a given prediction.

We will investigate two main model-agnostic interpretation methods: Local Interpretable Model-agnostic Explanations (LIME) and Counterfactual Explanations.

LIME approximates complex model predictions with simpler models to explain each prediction. By computing the model's predictions for a sample of data points around a given instance and weighting these points according to their proximity to the original instance, a simpler model, typically a linear regressor, can fit these weighted points. The coefficients of the simpler model provide an understandable depiction of the original model's decision-making procedure. The outcome provides a local explanation emphasising the features that contributed most to a given prediction, making it simpler to comprehend and verify the model's conclusion. LIME can be useful when it's not possible or too complicated to provide global explanations for how the model functions.

Counterfactual explanations address "what-if" scenarios and outline the minimum modification required to change the model's conclusion. A counterfactual explanation finds an alternate instance that would have obtained a different prediction given an instance and its prediction. It helps identify the factors most responsible for a particular prediction and clearly indicates how the model arrived at its decision. This method can be particularly useful when dealing with sensitive or high-risk applications where transparency and accuracy are critical. For instance, if a loan application is turned down, a counterfactual explanation can suggest that the loan would have been granted if the applicant had a little more money.

Interpretability is the ability to gain insight into the decision-making process of a machine learning model. It is particularly important in banking, healthcare, and criminal justice, where accountability and fairness are crucial. Counterfactual explanations can be incredibly helpful in these situations as they provide users with feedback that can help them understand and potentially change the causes of results.

Model-agnostic approaches are flexible means of interpreting the complex decision-making mechanisms of machine learning models. These strategies ensure more responsible and informed AI deployments by bridging the gap between sophisticated AI systems and human understanding. Examples of such approaches include answering "what-if" queries and offering local explanations for specific forecasts.

*Hybrid Models and Model Distillation*

Artificial intelligence and machine learning often require complex, multi-layered models to achieve the best performance. However, these models can be computationally expensive and difficult to understand, particularly in applications where clarity is essential. Hybrid models and model distillation techniques can help balance interpretability and complexity, allowing for easier knowledge transfer from complex models to simpler versions.

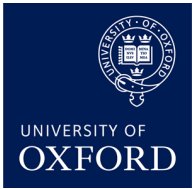
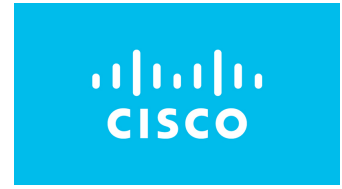

Hybrid models combine the benefits of simple and complex models to balance interpretability and complexity. Complex models, like deep neural networks, often sacrifice interpretability for accuracy, making their decision-making processes difficult to understand. Hybrid models aim to combine complex pattern recognition with interpretability by using a linear model for prediction and a deep learning architecture for feature extraction, for example. Hybrid models are particularly useful in industries like healthcare, where understanding the reasoning behind predictions is vital.

Model distillation is a process that involves teaching a less complex model to behave like a more complex one. The goal is to approximate the complex model's decision boundaries and general prediction skills without replicating its parameters. The process involves training a complex model on a dataset to a predetermined degree of proficiency and then training a simpler model using the complex model's predictions. During this process, the complex model's soft probabilities can provide the simpler model with additional information about the decision landscape. These techniques can help simplify complex models and make them more understandable while maintaining or improving performance.

Model distillation is a technique that offers several benefits compared to other, more complex models. Firstly, distilled models can be deployed on edge devices with limited resources because they are faster and require less memory. Additionally, they are typically easier to understand and debug than their more complicated counterparts, making them a more interpretable option. Despite their simplicity, distilled models can perform as well as teacher models if the distillation process is carried out correctly.

Distilled models are particularly useful when computing resources are limited but high performance is still required. This makes them ideal for real-time applications, mobile computing, and similar scenarios.

As AI spreads into new industries, the need for models that strike a balance between power and clarity will only increase. Model distillation and hybrid models are two cutting-edge approaches that ensure the AI systems of the future are transparent and powerful, promoting confidence and broad adoption.

### 4.2. The Significance of Dataset Transparency

Datasets are crucial in inspiring research, developing algorithms, and spurring inventions. However, the data's quality, comprehensiveness, and impartiality directly impact the reliability, equity, and integrity of analytical systems. Therefore, it is essential to ensure transparency in datasets.

The importance of representative and unbiased datasets cannot be overstated. Representative datasets produce models that can generalise to a wide range of populations, crucial for inclusivity and fairness. Biased datasets can unintentionally leave out or inaccurately portray segments of the population, leading to biased outcomes. Moreover, an impartial dataset preserves ethical norms by limiting the spread of stereotypes and prejudices. Biased conclusions derived from such datasets can have serious consequences for society, especially in vital areas like criminal justice, finance, and healthcare. Additionally, any innate prejudice in the data can reduce the effectiveness and accuracy of machine learning models.

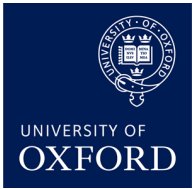
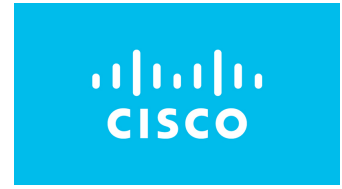

Various methods can be employed to identify and reduce dataset biases. The preliminary exploratory data analysis can identify distributional anomalies or disproportions in the dataset. Frequency analysis, cross-tabulation, and visualisation tools can aid this process. By assessing a model's behaviour under different conditions, bias impact assessment can identify any patterns of bias that may result from dataset biases. Techniques such as adversarial testing are used in this process. Data augmentation can help counteract underrepresented classes or circumstances by artificially increasing or changing the dataset, ensuring more equitable participation in all categories. Resampling can assist in reducing biases associated with class imbalance by striking a balance in the distribution of classes by either oversampling minority classes or undersampling majority ones.

Understanding the origin and documentation of datasets is crucial for determining their credibility. Having extensive metadata that explains the source, conditions, and data author can help assess the dataset's properties and any potential changes that might affect its reliability. This metadata should include detailed descriptions of the dataset's features, collection methods, and pre-processing activities. For datasets, versioning can be as beneficial as it is for software. It promotes transparency and reproducibility by ensuring that any changes, updates, or corrections to the dataset are systematically recorded.

To instil confidence in datasets and promote ethical data collection, it is important to record several details related to the data. These include ethical considerations, such as the potential impact of the data on individuals or groups, as well as consent procedures, which outline how the data was obtained and whether individuals were aware of its intended use. Additionally, it is crucial to identify any potential conflicts of interest that may have influenced the collection or use of the data. By taking these steps, data collectors and users can demonstrate their dedication to responsible and ethical practices, which can ultimately lead to greater trust and confidence in the data.

Transparency in datasets is fundamental to the moral, reliable, and effective operation of analytical systems; it is the cornerstone upon which fair and reliable artificial intelligence is built. Researchers, practitioners, and stakeholders are responsible for promoting transparency and accountability as the field of data science grows.

### 4.3. Accountability in the AI Lifecycle

Artificial Intelligence (AI) has made significant progress in various fields, from financial forecasting [49] to healthcare diagnostics [1], [50]–[52]. However, as AI algorithms become more advanced and powerful, it is essential to ensure that they are accountable throughout their lifespan. Accountability is both a moral responsibility and a technological necessity, ensuring that AI systems operate transparently, fairly, and in line with societal expectations.

Model developers play a critical role in establishing responsibility throughout the AI lifecycle. Their choices set the stage for AI systems' development and eventual operation, from selecting data sources to refining model architectures. Developers must prioritise ethics in the model development process, incorporating transparency, bias avoidance, privacy, and stakeholder engagement.

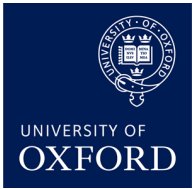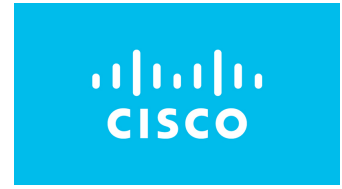

Transparency is crucial in creating models with interpretive decision-making processes that stakeholders can trust. Developers should be able to explain how the algorithm arrived at its decision, ensuring transparency and accountability. Additionally, developers must actively identify and address potential biases in model architectures and training data, preventing AI from unintentionally reinforcing or amplifying societal prejudices. They can use bias correction techniques or adjust the model's weighting to make it fairer.

Developers must also handle personal data appropriately, following data protection laws and using strategies such as differential privacy to safeguard specific data points while preserving the data's usefulness. Engaging with diverse stakeholders, such as ethicists and end-users, can provide better insight into ethical issues and best practices, ensuring that the model meets stakeholders' needs and expectations.

For models to be reliable and fair, they must undergo rigorous testing in different scenarios to ensure consistent performance and resistance to hostile attacks. Developers can use strategies like adversarial training to improve model resilience and make it more robust. They must also ensure that the model is fair by incorporating interventions that promote fairness, using fairness-related metrics to measure and reduce unjust prejudices. Developers must be aware of potential bias in the model and take steps to mitigate it. Regarding safety, it is crucial to prevent AI systems from accidentally causing harm or malfunctioning in unexpected situations. Safety protocols, error monitoring, and fail-safes can prevent adverse outcomes. Developers should consider potential safety risks and implement measures to mitigate them.

Model developers play a crucial and challenging role in the AI lifecycle. Their decisions impact interactions, outputs, and societal implications throughout the system's lifespan. Therefore, ensuring accountability requires continuous thought, improvement, and accountability instead of being a one-time event. The concept of accountability will continue to play a significant role in guiding AI's responsible and beneficial evolution as it gains global prominence.

### 4.4. Tech Companies and Implementers

Artificial Intelligence (AI) is becoming an increasingly important aspect of goods and services provided by tech businesses. However, with great power comes great responsibility. Since tech companies are the primary users of AI, they have a vital role to play in ensuring that their applications are morally sound, uphold societal norms, and safeguard individual privacy.

To achieve these goals, tech businesses should adopt a conscientious approach to integrating AI into their goods and services. This includes incorporating ethical considerations from the beginning of the design process using principles such as justice, transparency, and accountability to ensure that AI systems make fair decisions, are understandable to users, and have proper checks and balances. Privacy-by-design should also be embraced to ensure that data privacy is a fundamental aspect of developing AI systems, not optional.

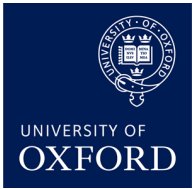
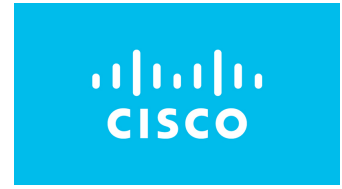

Another crucial aspect of conscientious AI integration is mitigating bias. Thorough bias testing should be performed on AI systems, particularly those used in decision-making processes. This includes prejudices based on socioeconomic class, gender, colour, and other factors. Businesses should invest in resources that identify and eliminate these prejudices.

Engagement of stakeholders is also essential in ensuring that AI systems are developed in a way that respects broader societal values and offers a range of perspectives. Tech companies should actively involve users, civil society, and subject matter experts in the development of AI.

In addition to conscientious AI integration, tech businesses should implement mechanisms for constant monitoring and feedback. This includes real-time observation of AI systems to ensure they function as intended and not cause unforeseen negative effects, even after deployment. Clear channels for users and other stakeholders to provide feedback on AI systems should be established. This feedback can be very helpful in identifying problems that need to be visible during the testing stage.

Artificial intelligence is an ever-evolving field, and therefore, products and services must be updated regularly to reflect the latest developments and ethical principles. Feedback should be used to make incremental changes, and organisations should issue transparency reports that provide information about their AI implementations, any issues identified, and the solutions implemented.

Finally, independent third-party audits can objectively assess whether a company's AI system installations are ethically sound and compliant with best practices. Companies should be open to these audits and follow the recommendations provided. By adopting a conscientious approach to integrating AI into their goods and services and implementing mechanisms for constant monitoring and feedback, tech businesses can uphold their responsibility to society and promote the ethical use of AI.

Artificial Intelligence (AI) has opened new horizons for digital firms, offering them unparalleled opportunities. However, the use of AI also requires a fresh commitment to ethics and accountability. IT firms must ensure that their AI services and products respect people's rights and benefit society. By incorporating ethical considerations from the very beginning, keeping a close eye on deployments, and being receptive to feedback, IT firms can ensure that their AI products are safe and ethical.

Red-teaming is a well-known cybersecurity technique businesses use to identify system vulnerabilities. The same technique can be applied to AI models. As AI systems become more integrated into critical financial services, healthcare, and transportation applications, they must be resilient to unforeseen events and malicious intent. This is where red-team testing comes in.

Businesses can use a structured approach to conduct red-team testing on newly developed AI models. The first step is to set objectives and describe the purpose of the exercise. This involves indicating the specific AI systems or models to be examined and the cases to be analysed. The objectives of the red-teaming exercise should be established, such as evaluating the model's ability to withstand hostile attacks, make ethical decisions, or manage specific edge circumstances.

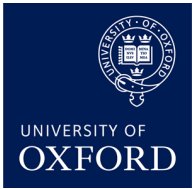
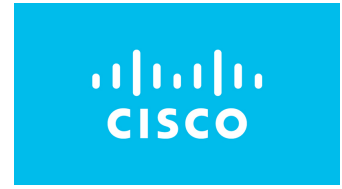

The second step is to assemble a red team comprising individuals with diverse expertise, such as ethical hackers, AI researchers, and domain specialists. External partners can also be brought in to provide an outside perspective, which can often identify problems internal teams may overlook due to familiarity bias.

The third step is to simulate attacks on the AI model. The red team can use various techniques such as adversarial attacks, data poisoning, ethical testing, and reverse engineering to identify any biases or weak areas in the model.

By following these steps, businesses can ensure that their AI models are robust and can withstand attacks. This will benefit companies and society as AI plays an increasingly important role in our lives.

Implementing a red-teaming process for AI model development involves several steps. The first step is to define the scope of the red-teaming process, which includes determining the specific objectives, constraints, and parameters. Next, a red team of experts in various fields, including cybersecurity, data science, and AI, should be built.

Once the red team is in place, they should develop a test plan to simulate possible rogue actors and test the AI model in a real environment. The red team should continuously test and probe the AI model in a real environment to identify any vulnerabilities or problems. Any issues found should be promptly reported to the development team to be addressed.

After the red-teaming process, comprehensive reports should be created that list all vulnerabilities found, their possible effects, and recommended fixes. These reports should be shared with all pertinent parties, including technical teams and upper management. The development team should address the vulnerabilities found to improve the security and resilience of the AI model. It is important to note that red-teaming should be a regular process to account for model changes and new risks.

By implementing red-teaming into the AI development lifecycle, AI models' robustness, fairness, and security can significantly increase. This is crucial to ensure that AI systems are resilient against rogue actors and unforeseen obstacles as they become more integral in various areas.

### 4.5. End-users and the General Public

Two crucial factors should be considered when discussing the intersection between end users, the public, and AI systems. Firstly, it is essential to educate people about the capabilities and limitations of these systems, and secondly, to encourage critical thinking and awareness of their use and influence. The primary objective is to ensure a society that is knowledgeable and actively involved in the creation and application of AI technology.

Teaching people about the benefits and drawbacks of AI systems can be achieved through various public education and outreach initiatives. This includes creating all-inclusive resources that produce understandable, non-jargonised teaching materials describing the capabilities and limitations of artificial intelligence. These resources can be available online, through educational institutions, and public libraries. Conducting workshops and seminars in collaboration with academic institutions, tech firms, or neighbourhood associations can also be helpful.

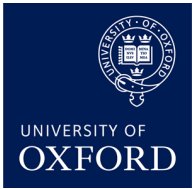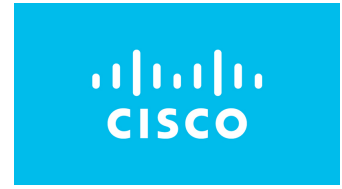

Participation via media is another way to educate the public about AI technologies. Information campaigns using social and conventional media can raise awareness of AI technologies, and visually appealing videos and infographics can be especially powerful. Encouraging the creation of documentaries and series that offer a nuanced perspective on AI technologies by examining their applications, ethical concerns, and use cases is also beneficial.

AI education should be incorporated into schools' curricula to help students grasp AI from an early age. More advanced courses exploring the inner workings of AI, including its promise and drawbacks, should be made available to individuals interested in the subject. Community centres may run talks or courses aimed at helping people comprehend artificial intelligence, and online platforms can offer accessible learning opportunities to people of all ages using MOOCs (Massive Open Online Courses) and other e-learning tools.

Encouraging awareness and critical thinking can be done by promoting forums and discussions concerning AI to address its implications for society. Encouraging media literacy also involves educating individuals to evaluate the accuracy of AI-related material in the media. Promoting examinations of the effects AI systems have on various industries and populations and bringing up ethical issues early in classroom discussions and public discourse can help recognise the social effects of AI.

We need to ensure that AI systems are developed and used in a way that benefits everyone and does not cause harm. One critical aspect of responsible AI development is engagement in AI governance. By actively involving the public in discussions and decision-making processes related to AI advancements and policies, we can create AI systems that are more democratically responsible. Advocacy and citizen groups must be established to represent public interests in venues for AI development.

In addition to public discussions and advocacy, we need to encourage businesses to be more transparent and responsible in using AI. This can be achieved through AI explainability, which makes it easier for the public to understand how AI systems operate. Furthermore, we need to create feedback channels that allow members of the public to report any problems or concerns about the AI systems they use.

By focusing on these areas, we can promote a society that benefits from AI while ensuring its responsible usage and development. It is crucial for the public to actively participate in defining the direction of AI technology rather than just consuming it as consumers. Together, we can shape a future where AI is used for the greater good.

## 5. Ethical AI design principles and guidelines

In the fifth chapter of this discourse, we will explore the importance of ethical AI design principles and guidelines. These principles and guidelines are crucial in directing the development and deployment of Artificial Intelligence (AI) towards a path that aligns with moral imperatives and societal values. In today's world where AI is becoming increasingly dominant, it is essential to establish robust ethical frameworks to ensure that AI systems are technologically advanced, socially responsible, and morally sound. In this chapter, we will delve into the core principles that form the ethical foundation of AI design, including transparency, accountability, fairness, and non-maleficence. The chapter delves into the guidelines proposed by global entities and think tanks, aimed at shaping AI into a force that upholds human dignity and societal welfare. By navigating through these principles and guidelines, the chapter provides a comprehensive blueprint for ethical AI design, which fosters trust and harmony between AI systems and the human society they intend to serve.

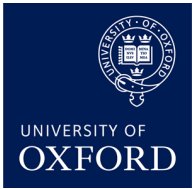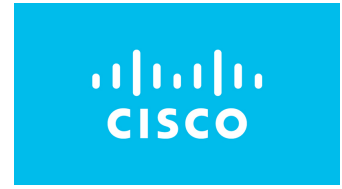

## 5.1. Regulatory and Legal Aspects

As AI technology advances rapidly, educating the public about its capabilities and limitations is essential. AI has the potential to revolutionise the way we live, work, and interact with each other. Still, promoting critical thinking and awareness about its effects and applications is equally important.

The primary objective of AI education is to ensure that people are well-informed and actively involved in developing and implementing AI technology. To achieve this, we need to create inclusive and accessible materials for public outreach and education programs that describe the advantages and drawbacks of artificial intelligence without relying on technical jargon.

Public libraries, educational institutions, and online platforms can be utilised to disseminate these materials to a broad audience. The materials can be tailored to different age groups and academic levels to ensure maximum impact.

Moreover, interactive workshops and seminars can be organised in collaboration with academic institutions, digital firms, or community associations to help people understand AI and its potential applications. These workshops can be tailored to different audiences, such as school children, university students, and professionals.

The workshops will allow the public to ask questions, share their concerns, and learn from experts in the field. They will also provide a platform for feedback, which will help improve the quality of the educational materials and the programs.

We can leverage social and traditional media to raise awareness of AI technologies. Information campaigns, infographics, and short films can be utilised to explain complex AI concepts easily. Additionally, we can promote the creation of documentaries that examine AI technologies, their applications, and the moral implications they raise, offering a sophisticated perspective on the field.

AI concepts can be included in education curricula early on, making it easier for students to understand the technology and its potential uses. Specialised courses can be offered to interested parties to explore AI's potential, drawbacks, and workings.

To promote awareness and critical thought, we should create forums and facilitate discussions about AI to address its implications for society. Encouraging media literacy is also crucial, empowering individuals to evaluate the AI-related content in the media. We can offer community courses and online platforms that make learning about AI accessible to people of all ages, thus promoting lifelong learning.

In summary, promoting AI education is essential to ensure that people are well-informed and actively involved in creating and implementing AI technology. By creating inclusive and accessible materials, organising workshops and seminars, leveraging social and traditional media, and including AI concepts in education curricula, we can raise awareness and promote critical thinking about AI technology.

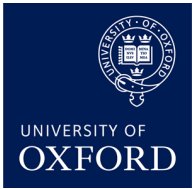
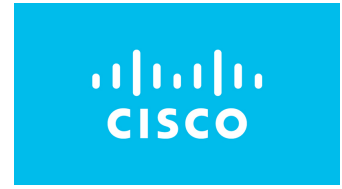

As AI becomes more pervasive, it is crucial to understand its impact on society. Conducting impact assessments on various industries and populations can provide valuable insights into how AI is affecting society and how it can be used responsibly.

Ethics is a crucial consideration when integrating AI into society. It is recommended that ethics topics be discussed in education and public discourse to promote a culture of moral technology use. This will ensure that AI is developed and used in a way that is responsible and respectful of all stakeholders.

Promoting the responsible development and application of AI requires involvement from the public in AI governance. Public input gatherings can provide feedback and insights from diverse stakeholders. Citizen and advocacy groups can also represent the public in venues for AI development, ensuring that AI is developed and applied ethically and equitably.

Accountability and transparency mechanisms are essential to promoting the responsible use of AI. AI explainability can encourage businesses to make their AI systems transparent and intelligible to the public. Feedback channels should also be created to allow the public to voice any concerns regarding AI systems. This will help identify issues and work collaboratively to address them.

As AI technologies become more integrated into society, it is increasingly important to establish strong and uniform regulatory frameworks to ensure that AI is used responsibly and ethically. Developing international standards and guidelines for global regulatory frameworks on AI ethics is crucial in achieving this. The OECD Guidelines [53] for AI and UNESCO's Ethics of AI Recommendation emphasise the importance of ensuring that AI respects human rights, democratic ideals, and the rule of law.

Many countries are developing domestic and international laws to promote the responsible development and application of AI. For example, the European Commission has proposed the Artificial Intelligence Act, a crucial step toward establishing legal guidelines for the reliable application of AI. Similarly, Singapore has created frameworks like the Model AI Governance Framework to help organisations use AI ethically. In the US, recommendations are being enforced to promote AI innovation and the development of moral AI practices.

By focusing on these areas, we can promote a society that benefits from AI and actively participates in its responsible development and application. The public needs to have an active and informed role in determining the direction of AI technology rather than simply being passive consumers.

The Global Partnership on AI (GPAI) is a collaboration of experts from various fields committed to the responsible development and application of AI. The organisation aims to promote international research and policy discussions on AI, to achieve a more ethical and transparent AI landscape. Alongside GPAI, the AI Ethics and Governance Alliance works towards similar objectives.

The impact of GDPR regulations on AI decision-making systems, specifically regarding processing personal data, profiling, and automated decision-making, cannot be overstated. GDPR guidelines require AI systems to collect only the minimum amount of data necessary for the intended use, maintain detailed records of all data processing operations, and ensure transparency and accountability. However, there is an ongoing debate regarding the delicate balance between AI innovation and data privacy regulations.

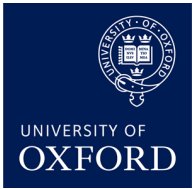
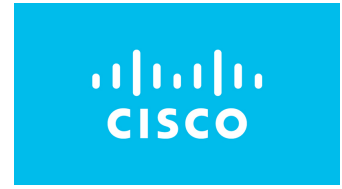

Furthermore, global AI services need help in dealing with multiple data protection environments, and there is a need for standardised laws worldwide without restricting creativity. To maintain public trust in technology, ethical AI design must work hand in hand with developing legal requirements, thereby promoting a more responsible use of AI.

## 6. The role of AI in decision-making: ethical implications and potential consequences

Advanced machine learning techniques are increasingly utilised in decision-making, ranging from routine product recommendations to critical healthcare, banking, and law enforcement decisions. AI's ability to process vast amounts of data and detect patterns enhances efficiency, objectivity, and precision.

The level of autonomy granted to AI systems is a debated issue, particularly when decisions have significant implications for people or communities. A risk-based assessment is often used to determine the level of autonomy required, with higher-risk domains requiring more human oversight. Ethical considerations like human dignity and moral responsibility are important when balancing human intervention with AI autonomy.

Reliability and resilience are crucial in ensuring that AI systems are reliable and robust, and they should be aware of their limitations and seek human assistance if necessary. Ethical frameworks and regulations must evolve as AI technologies evolve to ensure that AI systems serve the public good while respecting individual rights and social values. Multi-stakeholder governance structures and public participation can help navigate these complex moral environments.

## 7. Establishing responsible AI governance and oversight

Effective governance and oversight of AI is a complex and multi-faceted issue requiring input from diverse stakeholders, including lawmakers, industry experts, technologists, ethicists, and the public. The overarching goal is to ensure that AI is developed and implemented responsibly, ethically, transparently, and human-centred that aligns with societal values and human rights.

A few recommended steps should be taken to establish responsible AI governance and oversight. First, it is crucial to create ethical structures that define the moral principles of AI, such as justice, transparency, fairness, privacy, and security. Sociologists and ethicists should collaborate to develop these norms, considering various moral and cultural viewpoints to ensure inclusivity and diversity.

Second, policy and regulation ensure that AI systems are developed and used responsibly. Guidelines should be established for appropriate AI applications, and accountability systems should be in place to ensure that regulations are followed. Regulations should be flexible enough to evolve with the rapidly changing landscape of AI technology while striking a balance between innovation and accountability to ensure their effectiveness.

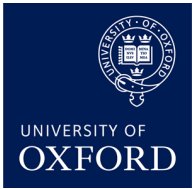
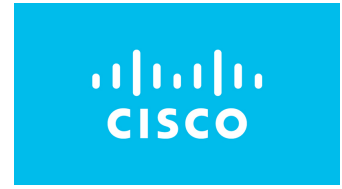

Overall, the governance and oversight of AI must be approached with care and consideration to ensure that AI is developed and used in ways that benefit society while minimising potential harm.

The responsible development and application of AI is a complex issue that requires careful consideration of multiple factors. One critical element is the establishment of standards and certification procedures to ensure that AI is developed and applied ethically. By collaborating with global standardisation organisations, such as the International Organisation for Standardisation (ISO), the industry can create guidelines that cover various aspects of AI, such as data management, algorithmic transparency, explainability, and fairness.

In addition to standards and certification, institutional oversight is also essential. Independent organisations with the power to supervise AI applications and enforce rules must be established to provide this oversight. These organisations should be independent, accountable to the public, and subject to regular audits. Strong institutional oversight can prevent abuses and ensure that AI is developed and used responsibly.

However, responsible AI governance and oversight go beyond standards, certification, and institutional oversight. It also requires the development of ethical structures, policies, and regulations to guide the development and use of AI. These structures, policies, and principles should be based on a collaborative approach involving multiple stakeholders, including government, industry, academia, and civil society.

Moreover, the organisations responsible for AI governance and oversight should have the expertise to assess AI systems for conformity to moral standards and legal requirements without compromising privacy or security. They should be able to evaluate the potential benefits and risks associated with AI and identify areas where AI can be used responsibly and should not be used.

Responsible AI governance and oversight require a collaborative approach involving multiple stakeholders, ethical structures, policies and regulations, standards and certification procedures, and institutional management. By establishing these elements, we can ensure that AI benefits society and is developed and used ethically.

There are several key considerations to remember to ensure that AI systems are developed and governed responsibly and ethically. First and foremost, transparency is crucial. This means providing detailed information about how AI systems are designed, trained, and tested and how they make decisions and handle errors. By being transparent and open about the algorithms and decision-making processes used, we can promote fair and accountable AI systems that build trust and confidence among the public.

Another important aspect of AI governance is public education and engagement. It's crucial to involve a broad range of perspectives and interests, including those of the public, in discussions about the implications and impact of AI on society. This can be achieved through public consultations, online forums, and other participatory processes. Providing accessible and understandable information about AI, including its benefits, risks, and limitations, can help promote a deeper understanding of AI and its impact.

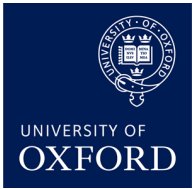
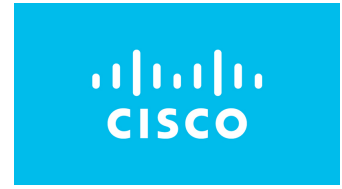

When it comes to analysing the potential and limitations of AI, multidisciplinary studies are key. These studies should involve computer science, law, political science, ethics, and social sciences experts. By analysing the impact of AI on governance efficiency, effectiveness, and legitimacy, we can ensure that AI is developed and governed in a way aligned with public values and interests, maximising its benefits while minimising its risks.

Finally, constant monitoring and adaptation ensure that AI systems are developed and governed responsibly. AI systems are complex and can have unintended consequences, so monitoring their performance and impact continuously and adapting governance frameworks accordingly is essential [54], [55]. By setting up procedures for ongoing monitoring, evaluation, and feedback, we can ensure that the governance structure remains flexible and adaptive, able to respond quickly to new data, technological advancements, and changing public demands. Ultimately, by constantly monitoring and adapting AI systems, we can ensure that they are developed and governed in a way that is responsible and ethical and that all share their benefits. It is possible to maximise AI's benefits while lowering its hazards by integrating ethical standards into every stage of the technology's lifespan and ensuring a strong structure of governance and monitoring is in place. This calls for alertness, vision, and a steadfast dedication to the ideals of democratic control and human-centred technological progress.

## 8. AI in sensitive domains: healthcare, finance, criminal justice, defence, and human resources

Deploying artificial intelligence (AI) in sensitive fields such as healthcare, finance, criminal justice, defence, and human resources presents unique challenges and moral dilemmas that require careful attention.

In healthcare, AI has the potential to revolutionise patient care by enabling early diagnosis, personalised therapy, and optimised treatment. However, there are concerns regarding patient privacy, data consent, decision-making bias, and the accuracy of AI-powered diagnoses. To address these concerns, healthcare laws such as Europe's General Data Protection Regulation (GDPR) and the USA's Health Insurance Portability and Accountability Act (HIPAA) [56] must be complied with to protect patient data.

AI can enhance automated trading, fraud detection, and risk assessment in finance. However, there is a risk of algorithmic bias, unclear financial advice, and market instability caused by AI. Financial AI systems must comply with legal frameworks such as Europe's Markets in Financial Instruments Directive (MiFID II) to mitigate these risks, ensuring transparency and consumer protection.

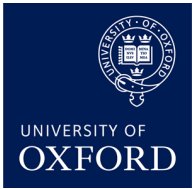
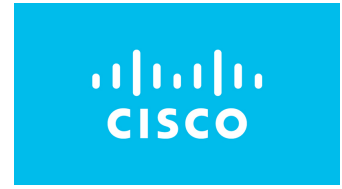

In criminal justice, AI applications like predictive policing and risk assessment tools are controversial as they may infringe on civil liberties and exacerbate existing biases. To prevent these risks, governing bodies must scrutinise the use of these technologies to ensure compliance with human rights legislation and prevent systemic discrimination.

In defence, AI is used for autonomous weapon systems, logistics, and monitoring. This raises serious ethical questions, particularly with the development of lethal autonomous weapon systems (LAWS). It is necessary to regulate such technology in accordance with international humanitarian law and the United Nations Convention on Certain Conventional Weapons (CCW) to ensure compliance with ethical standards and prevent the misuse of AI in warfare.

AI-driven recruitment, performance monitoring, and management solutions can greatly enhance the efficiency of HR processes. However, it is important to address privacy concerns, potential biases in recruitment methods, and the need for transparency in decisions that affect employees' employment. Regulations such as the GDPR provide some control over automated decision-making and profiling.

It is essential to consider the specific legal circumstances, ethical considerations, and social implications associated with the implementation of AI in various industries to ensure that AI systems are appropriately deployed. AI systems in these areas must be equitable, transparent, and auditable. Effective accountability and resolution mechanisms must be effective, and all stakeholders must work together to establish trust in using AI. To create AI systems that are not only technologically advanced but also socially responsible and ethically sound, interdisciplinary collaborations between computer scientists, ethicists, legal experts, and domain specialists are critical.

## 9. Discussion on engaging stakeholders: fostering dialogue and collaboration between developers, users, and affected communities.

Stakeholder engagement is critical to developing and deploying Artificial Intelligence (AI) technologies. Ensuring that AI technologies are created and implemented ethically, responsibly, and beneficially for society is crucial. AI systems profoundly impact various aspects of human life, such as healthcare, education, finance, transportation, and security. Therefore, it is vital to establish communication and collaboration among developers, users, and affected communities to understand and address the ethical concerns that new technologies bring.

To engage stakeholders effectively, it is necessary to provide platforms and methods that enable open communication and collaboration. These platforms can include public consultations, stakeholder seminars, and cross-sector cooperation. The aim of these platforms should be to identify concerns, co-create solutions, build trust, and promote literacy.

Identifying concerns involves gathering input from various stakeholders to identify privacy, bias, accountability, transparency, and other ethical issues. This involves engaging with diverse groups such as policymakers, industry leaders, civil society organisations, academics, and affected communities. Co-creating solutions involve collaborating with stakeholders to develop solutions that address their issues, such as defining ethical guidelines, best practices, and governance frameworks. The solutions should reflect the values and needs of all stakeholders and be flexible enough to adapt to changing circumstances. Building trust involves developing trust among stakeholders by demonstrating a commitment to resolving ethical considerations and ensuring that AI systems align with social values and norms. This can be achieved by establishing clear accountability mechanisms, transparency standards, and ethical codes of conduct. Promoting literacy involves increasing AI literacy among all stakeholders to enable more informed and productive discussions regarding the ethical issues of AI. This includes educating stakeholders about the benefits and risks of AI, the limitations of AI technologies, and the ethical principles that guide AI development and deployment.

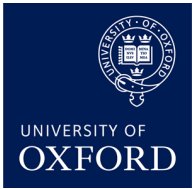
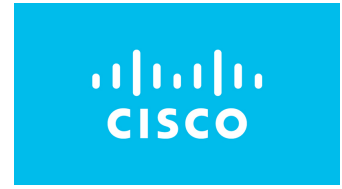

It is crucial to anticipate how ethical concerns may emerge as AI technologies advance. As AI systems become more sophisticated and their decision-making processes opaquer, ensuring transparency and explainability will become increasingly challenging. This requires developing new techniques and technologies that enable AI systems to be more transparent, interpretable, and accountable. The debate about the degree of autonomy given to AI systems and the amount of control retained by humans will intensify, particularly as autonomous systems become more prevalent. This requires developing new governance models and regulatory frameworks that balance the benefits of AI with the risks and ensure that humans remain in control of critical decisions.

## 10. Conclusion

Artificial intelligence (AI) has the potential to significantly impact employment, social equity, and economic systems in ways that require careful ethical analysis and aggressive legislative measures to mitigate negative consequences. This means that the implications of AI in different industries, such as healthcare, finance, and transportation, must be carefully considered.

Due to the global nature of AI technology, global collaboration must be fostered to establish standards and regulatory frameworks that transcend national boundaries. This includes the establishment of ethical guidelines that AI researchers and developers worldwide should follow.

To address emergent ethical concerns with AI, future research must focus on several recommendations. Firstly, ethical considerations must be integrated into the design phase of AI systems and not treated as an afterthought. This is known as "Ethics by Design" and involves incorporating ethical standards during the development phase of AI systems to ensure that the technology aligns with ethical principles.

Secondly, interdisciplinary research that combines AI, ethics, law, social science, and other relevant domains should be promoted to produce well-rounded solutions to ethical dilemmas. This requires the participation of experts from different fields to identify and address ethical issues.

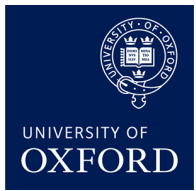
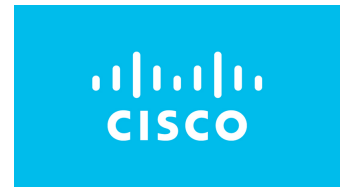

Thirdly, regulatory frameworks must be dynamic and adaptive to keep pace with the rapid evolution of AI technologies. This means that regulatory frameworks must be flexible enough to accommodate changes in AI technology while ensuring ethical standards are maintained.

Fourthly, empirical research should be conducted to understand the real-world implications of AI systems on individuals and society, which can then inform ethical principles and policies. This means that empirical data must be collected to understand how AI affects people in different contexts.

Finally, risk assessment procedures should be improved to better analyse the ethical hazards associated with AI applications, considering the intricacies of varied societal situations. Risk assessment methodologies must be developed to identify ethical risks in different contexts.

By proactively anticipating future ethical difficulties and employing a collaborative, multidisciplinary approach, the AI community can steer the development of AI technology towards outcomes that align with agreed ethical ideals and societal goals. This requires ongoing awareness of ethical principles, innovation, and a commitment to ethical standards as the capabilities and applications of AI continue to expand.

https://web.archive.org/web/20170312045558/http://nvlpubs.nist.gov/nistpubs/FIPS/NIST.FIPS.197.pdf

[23] C. S. Thirumalai, S. Budugutta, and C. Thirumalai, "Public key encryption for SAFE transfer of one time password heuristic prediction of olympic medals using machine learning View project Extreme Machine Learning View project Public Key Encryption for SAFE Transfer of One Time Password", Accessed: Sep. 22, 2023. [Online]. Available: https://www.researchgate.net/publication/323277440

[24] Q. Zhang, S. Jia, B. Chang, and B. Chen, "Ensuring data confidentiality via plausibly deniable encryption and secure deletion – a survey," *Cybersecurity*, vol. 1, no. 1, pp. 1–20, Dec. 2018, doi: 10.1186/s42400-018-0005-8.

[25] GDPR, "What is GDPR, the EU's new data protection law? - GDPR.eu." Accessed: Jul. 07, 2023. [Online]. Available: https://gdpr.eu/what-is-gdpr/

[26] ICO, "Information Commissioner's Office (ICO): The UK GDPR," UK GDPR guidance and resources. Accessed: Jul. 08, 2023. [Online]. Available: https://ico.org.uk/for-organisations/uk-gdpr-guidance-and-resources/lawful-basis/a-guide-to-lawful-basis/lawful-basis-for-processing/consent/

[27] T. Kovanen, J. Pöyhönen, and M. Lehto, "Cyber-Threat Analysis in the Remote Pilotage System," in *ECCWS 2021 20th European Conference on Cyber Warfare and Security*, Academic Conferences Inter Ltd, 2021, p. 221.

[28] D. Golovin, B. Solnik, S. Moitra, G. Kochanski, J. Karro, and D. Sculley, "Google vizier: A service for black-box optimization," in *Proceedings of the ACM SIGKDD International Conference on Knowledge Discovery and Data Mining*, New York, NY, USA: Association for Computing Machinery, Aug. 2017, pp. 1487–1496. doi: 10.1145/3097983.3098043.

[29] J. Waring, C. Lindvall, and R. Umeton, "Automated machine learning: Review of the state-of-the-art and opportunities for healthcare," *Artificial Intelligence in Medicine*, vol. 104. Elsevier B.V., p. 101822, Apr. 01, 2020. doi: 10.1016/j.artmed.2020.101822.

[30] I. Drori *et al.*, "AlphaD3M: Machine learning pipeline synthesis," in *AutoML Workshop at ICML*, 2018.

[31] Y. Bengio, A. Courville, and P. Vincent, "Representation learning: A review and new perspectives," *IEEE Trans Pattern Anal Mach Intell*, vol. 35, no. 8, pp. 1798–1828, 2013, doi: 10.1109/TPAMI.2013.50.

[32] F. Mohr, M. Wever, and E. Hüllermeier, "ML-Plan: Automated machine learning via hierarchical planning," *Mach Learn*, vol. 107, no. 8–10, pp. 1495–1515, Sep. 2018, doi: 10.1007/s10994-018-5735-z.

[33] N. Fusi, R. Sheth, and M. Elibol, "Probabilistic matrix factorization for automated machine learning," *Adv Neural Inf Process Syst*, vol. 31, pp. 3348–3357, 2018.

[34] D. Dasgupta, Z. Akhtar, and S. Sen, "Machine learning in cybersecurity: a comprehensive survey," *The Journal of Defense Modeling and Simulation:*